\documentstyle[12pt]{article}
\newcommand{\journal}[4]{{\em #1~}#2\,(19#3)\,#4.}

\newcommand{\ijmp}{\journal {Int. J. Mod. Phys.}}

\newcommand{\pr}{\journal {Phys. Rev.}}

\newcommand{\cmp}{\journal {Comm. Math. Phys.}}

\newcommand{\zp}{\journal {Z. Phys.}}
\newcommand{\np}{\journal {Nucl. Phys.}}
\newcommand{\pl}{\journal {Phys. Lett.}}

\newcommand{\prep}{\journal {Phys. Reports}}

\newcommand{\nc}{\journal {Nuovo Cim.}}

\makeatletter
\@addtoreset{equation}{section}
\makeatother

\def\Lp{\displaystyle{\biggl(}}
\def\Rp{\displaystyle{\biggr)}}

\newcommand{\lp}{\left(}\newcommand{\rp}{\right)}

\newcommand{\G}{\Gamma}
\newcommand{\D}{\Delta}
\renewcommand{\a}{\alpha}

\renewcommand{\d}{\delta}
\newcommand{\e}{\varepsilon}

\newcommand{\m}{\mu}
\newcommand{\n}{\nu}

 \renewcommand{\O}{\Omega}

\newcommand{\s}{\sigma} \renewcommand{\S}{\Sigma}

\newcommand{\Oh}{\widehat\Omega}
\renewcommand{\t}{\tau}

\newcommand{\BS}{{\cal B}_{\Sigma_m}}

\newcommand{\FF}{{\cal F}}

\newcommand{\SS}{{\cal S}}

\newcommand{\complex}{{\kern .1em {\raise .47ex
\hbox {$\scriptscriptstyle |$}}
    \kern -.4em {\rm C}}}
\newcommand{\real}{{{\rm I} \kern -.19em {\rm R}}}
\newcommand{\rational}{{\kern .1em {\raise .47ex
\hbox{$\scripscriptstyle |$}}
    \kern -.35em {\rm Q}}}
\renewcommand{\natural}{{\vrule height 1.6ex width
.05em depth 0ex \kern -.35em {\rm N}}}

\newcommand{\cb}{{\bar c}}
\newcommand{\half}{\frac 1 2}
\newcommand{\pa}{\partial}
\newcommand{\pad}[2]{{\frac{\partial #1}{\partial #2}}}
\newcommand{\fud}[2]  {{\displaystyle{\frac{\delta #1}{\delta #2}}}}

\newcommand{\ie}{{{\em i.e.}\ }}

\newcommand{\sla}{\raise.15ex\hbox{$/$}\kern -.57em}

\newcommand{\twiddle}{\lower.9ex\rlap{$\kern -.1em\scriptstyle\sim$}}

\newcommand{\th}{{\hat\tau}}

\renewcommand{\=}{=&} 
\newcommand{\equ}[1]{(\ref{#1})}
\newcommand{\Sm}{{\S_{m}}}
\newcommand{\eq}{\begin{equation}}
\newcommand{\eqn}[1]{\label{#1}\end{equation}}
\newcommand{\eea}{\end{eqnarray}}
\newcommand{\eqa}{\eq\ba{rl}}
\newcommand{\eqan}[1]{\ea\label{#1}\end{equation}}
\newcommand{\ba}{\begin{array}}
\newcommand{\ea}{\end{array}}
\newcommand{\eqac}{\begin{equation}\begin{array}{rcl}}
\newcommand{\eqacn}[1]{\end{array}\label{#1}\end{equation}}

\renewcommand{\pad}[2]{{\displaystyle{\frac{\partial #1}{\partial #2}}}}

\def\intx{\displaystyle{\int d^4 \! x \, }}

\begin{document}
\def\ftoday{{\sl  \number\day \space\ifcase\month
\or Janvier\or F\'evrier\or Mars\or avril\or Mai
\or Juin\or Juillet\or Ao\^ut\or Septembre\or Octobre
\or Novembre \or D\'ecembre\fi
\space  \number\year}}



\titlepage

\begin{center}

{\huge  Infrared Regularization of Yang--Mills Theories }

\vspace{2cm}

{\Large Alberto Blasi}

{\it Dipartimento di Fisica -- Universit\`a di Genova\\
via Dodecaneso 33 -- I-16146 Genova\\Italy}

and

\vspace{1cm}

{\Large Nicola Maggiore}\footnote{Supported in part
by the Swiss National Science Foundation. On leave of absence from
Dipartimento di Fisica -- Universit\`a di Genova, Italy}

{\it D\'epartement de Physique Th\'eorique --
     Universit\'e de Gen\`eve\\24, quai E. Ansermet -- CH-1211 Gen\`eve
     4\\Switzerland}

\end{center}

\vspace{2.5cm}

\begin{center}
\bf ABSTRACT
\end{center}

{\it We introduce an infrared regulator in Yang--Mills theories under
the form of a mass term for the nonabelian fields. We show that the
resulting action, built in a covariant linear gauge, is multiplicatively
renormalizable by proving the validity at all orders of the Slavnov
identity defining the theory.}

\vfill
\noindent
hep-th/9511068\\
UGVA-DPT-96-03-236 

\newpage

\section{Introduction}

It is a commonly accepted statement that the Higgs mechanism
constitutes the only way to give masses to the gauge bosons in the framework
of a local, renormalizable and unitary theory. On the other hand, the Higgs
boson, \ie the particle whose existence is entailed by the Yang-Mills (YM)
theory,
has not been discovered yet. This negative experimental fact renders meaningful
all efforts towards alternative theories describing massive YM
fields. The task is formidable, and despite the many trials attempted in the
last thirty years, no alternative theory to the Yang--Mills--Higgs model has
been proposed in order to give mass to the gauge bosons and
to fulfill the three necessary requirements of
\begin{enumerate}
\item locality
\item renormalizability
\item unitarity.
\end{enumerate}
In particular, the issue of unitarity for massive YM theory can be
satisfied only with a spontaneous symmetry breaking mechanism.
This old and somehow dormant subject is now knowing a kind of second youth.
For instance, much discussion arose again on the old Curci-Ferrari (CF)
model~\cite{cf}, due to some very recent contributions~\cite{pvn,peri}.
In this note, we first recall the basic features of the CF theory
and we explain the reasons why it cannot be considered as alternative
to the Yang-Mills-Higgs model.
Then, we propose a very simple way to give masses to the YM
fields by means of a local and renormalizable model, without addressing
the problem of its unitarity, our aim being that of giving a BRS invariant
infrared regularization for the nonabelian fields. In fact
a completely satisfactory regularization for YM theories
at low momenta is not currently available;
it is indeed known  that giving masses to the YM vector bosons through
the Higgs mechanism,may destroy the asymptotic freedom. This is basically
due to the fact that the contribution of the scalars tends to make
 positive the slope of the beta function in YM theories. Together with the
non--discovery of the Higgs particle, this fact constitutes another good
reason for avoiding scalars altogether.

\section{The Curci-Ferrari model}

The most interesting local and renormalizable massive non-abelian gauge model
not involving scalars is still represented
by the CF model~\cite{cf}, whose classical action
contains the mass term
\eq
S_{mass}^{CF}=\intx\Lp \frac{m^2}{2}A^2 + \a m^2\cb^ac^a\Rp\ ,
\eqn{cfmass}
where $(\cb)c$ is the (anti)ghost and $\a$ is the gauge parameter.
Once the gauge is fixed, the propagator for the YM fields
reads
\eq
\D_{\m\n}^{ab}=\frac{\d^{ab}}{k^2-m^2}
\lp g_{\m\n}-\frac{k_\m k_\n}{k^2} \rp
+\a\frac{k_\m k_\n}{k^2} \frac{\d^{ab}}{k^2-\a m^2}\ .
\eqn{prop}
An unphysical pole appears at $k^2=\a m^2$. The fact that, due to the
particular mass term~\equ{cfmass}, an identical pole is present in the ghost
propagator, justified the hope that the CF model could be unitary.

The gauge--fixing term of the CF action is
\eq
S_{gf}^{CF} = \intx \Lp b^a\pa A^a +\cb^a\pa^\m(D_\m c)^a
+\half \a b^2
 - \half \a f^{abc}b^a\cb^bc^c-\frac{1}{8}
\a f^{abc}\cb^b\cb^cf^{cmn}c^mc^n \Rp\ ,
\eqn{cfgf}
where $b(x)$ is the Lagrange multiplier usually introduced to enforce
the gauge condition. The total CF action
\eq
S^{CF} = -\frac{1}{4g^2}\intx F^2 + S^{CF}_{gf}
+S^{CF}_{mass}
\eqn{cfaction}
is invariant under the set of field transformations
\eqa
s^{CF} A^a_\m \= -(D_\m c)^a \nonumber \\
s^{CF}c^a \= \half f^{abc}c^bc^c \nonumber \\
s^{CF}\cb^a \= b^a \\
s^{CF} b^a \= -m^2 c^a \nonumber\ .
\eqan{cfbrs}
The CF model is affected by a few problems, here we would like to
stress the most serious ones~:
\begin{description}
\item[Non-linearity of the gauge condition]
The equation of motion of the Lagrange multiplier is nonlinear
\eq
\fud{S^{CF}}{b^a} = \pa A^a +\a b^a -\half\a f^{abc}\cb^bc^c
\eqn{cfgaugecond}
and therefore the gauge condition defined by~\equ{cfgaugecond} holds only
classically, since it cannot be implemented at the quantum level, contrary
to what happens in the ordinary, massless, case. As a consequence,
the multiplier cannot be eliminated from the quantum action, and
an additional symmetry must be used to guarantee the renormalizability of the
model~\cite{delsor}
\eqa
\d\cb^a \= c^a \nonumber\\
\d b^a \= \half f^{abc} c^bc^c \\
\d A^a_\m \= \d c^a =0 \nonumber
\eqan{delta}
However, the most unpleasant implication of the non linear gauge-fixing
condition~\equ{cfgaugecond} is that the hypersurface
crossing the gauge orbits in order to choose a representative for each of them,
is not defined for the quantum theory. This fact, which is related
to the presence in the CF action of a quartic term in the ghosts, leads to
a weakness of the very concept of gauge fixing for nonlinear choices
like~\equ{cfgaugecond}.
\item[Non-unitarity]
The CF model is invariant under the transformations~\equ{cfbrs}, which,
being non-nilpotent
\eq
(s^{CF})^2 = -m^2\d\ ,
\eqn{cfalg}
cannot even be classified as of the BRS type.
The lack of nilpotency of the pseudo-BRS operator~\equ{cfbrs} did
not prevent from showing that the model is renormalizable~\cite{cf1}
by means of five multiplicative renormalization constants~\cite{delsor}.
 In general, the physical states
coincide with the cohomology classes of the nilpotent BRS operator defining
the theory. The definition of ``physical space'' for a theory described by
a non-nilpotent BRS operator is not clear, and it is not at all
surprising that the non-nilpotency of the symmetry describing the theory
is the central point of the proofs on non-unitarity of the CF
model~\cite{cf1,ojima,bau,del,pvn}, the first of which was given
by Curci and Ferrari themselves~\cite{cf1}. Later,
Ojima~\cite{ojima} explicitly found a state with
negative norm between the ``would be'' physical states of the CF model,
thus undoubtedly concluding about the lack of unitarity. Ojima's proof was
very recently improved in~\cite{pvn}, where a whole class of ``physical
states'' having negative norm has been found.
\item[Mass dependence]
Finally, whatever the observables of the CF model are, the control
of their dependence on the parameter~$m^2$, considered as an infrared
regulator and not as a  physical mass, is difficult, and it has not been
accomplished.
\end{description}

\section{Infrared Regularizated Theory}

In this section we introduce a mass for the vector bosons,
relaxing the condition of unitarity. In other words, here the mass must be
considered as an infrared regulator, which eventually will be set equal to
zero.
The objective is to write a local action for massive YM fields, which is
invariant under a set of nilpotent BRS transformations, whose cohomology
classes
-- and hence the physical operators -- are the same as in the massless case.

We start from the massless gauge-fixed YM action
\eq
S=-\frac{1}{4g^2}\intx F^2 + s \intx\Lp\cb^a\pa A^a
+\frac{\a}{2}\cb^ab^a\Rp\ ,
\eqn{S}
which is invariant under the usual nilpotent BRS field transformations
\eqa
s A^a_\m \= -(D_\m c)^a \nonumber \\
s c^a \= \half f^{abc}c^bc^c \nonumber \\
s \cb^a \= b^a \\
s b^a \= 0 \nonumber\ .
\eqan{brs}
Then, we introduce two external sources $\{\s(x),\t(x)\}$, organized in a BRS
doublet
\eqa
s \s \= \t \nonumber\\
s \t \= 0\ ,
\eqan{doublet}
and we add to the action~\equ{S} the cocycle
\eq
s\intx\s  \frac{A^2}{2}\ .
\eqn{trick}
Performing a shift in the source $\t(x)$
\eq
\t(x)\rightarrow\th(x)\equiv\t(x) +m^2\ ,
\eqn{shift}
the classical action
\eqa
S_{mass} \= -\frac{1}{4g^2}\intx F^2
 + s \intx\Lp\cb^a\pa A^a
 +\frac{\a}{2}\cb^ab^a +\s\frac{A^2}{2}\Rp
\nonumber\\
\= \intx \Lp  -\frac{1}{4g^2}F^2 + b^a\pa A^a +
 \cb^a\pa^\m(D_\m c)^a + \frac{\a}{2}b^2 + (\t +m^2) \frac{A^2}{2}
 +\s A^a_\m\pa^\m c^a \Rp
\eqan{Smass}
acquires an infrared regulator under the form of a mass term for the YM fields.

The action $S_{mass}$ is invariant under the nilpotent BRS
transformations~\equ{brs}, enlarged by the doublet~\equ{doublet}.
Moreover, as in the massless case, the gauge condition
is implemented by the following $linear$ field equation of the Lagrange
multiplier
\eq
\fud{S_{mass}}{b^a}=\pa A^a +\a b^a\ .
\eqn{gaugecond}
In the Landau gauge, \ie for $\a=0$, the action $S_{mass}$ satisfies two
additional constraints~: the ghost equation of the Landau gauge~\cite{bps}
\eq
\FF^aS_{mass}=\intx\Lp\fud{}{c^a}+f^{abc}\cb^b\fud{}{b^c}\Rp S_{mass} =0\ ,
\eqn{bps}
and the following identity
\eq
WS_{mass}=\intx\Lp\fud{}{\s}+c^a\fud{}{b^a}\Rp S_{mass}=0\ .
\eqn{w}
The Slavnov identity describing the theory is
\eq
\SS(\Sm)=\intx \Lp
\fud{\Sm}{\O^{a\m}}\fud{\Sm}{A^a_\m}
+\fud{\Sm}{L^a}\fud{\Sm}{c^a}
+\cb^a\fud{\Sm}{b^a}
+\th\fud{\Sm}{\s} \Rp =0\ .
\eqn{slavnov}
In~\equ{slavnov}, $\Sm$ is the classical action $S_{mass}$, increased
by a source term
\eq
\Sm=S_{mass}+S_{ext}\ ,
\eqn{sm}
introduced to define the composite operators constituted by the nonlinear
BRS transformations of the quantum fields
\eq
S_{ext} = \intx \Lp
\O^{a\m}sA^a_\m + L^a sc^a \Rp\ ,
\eqn{sext}
by means of two external sources $\O(x)$ and $L(x)$.

The proof of the renormalizability of the theory develops according to
straightforward lines~\cite{psbk}. Because of the presence of the massive
parameter~$m^2$, one must distinguish between ultraviolet and infrared
dimensions of the fields, according to the large- and small- momentum behaviour
of the propagators. The result is listed in the table, together with the
Faddeev-Popov assignments.
\begin{center}
\begin{tabular}{|l|r|r|r|r|r|r|r|r|}\hline
 & $A_\mu^a$ & $c^a$ & $\cb^a$ & $b^a$ & $\s$ & $\t$ & $\O^{a\m}$ &
                   $L^a$
\\ \hline
$d_{uv}$ & $1$ & $0$  & $2$ & $2$ & $2$ & $2$ & $3$ & $4$
\\ \hline
$d_{ir}$ & $2$ & $0$  & $2$ & $2$ & $\geq 2$ & $\geq 2$ & $\geq 3$ & $\geq 4$
\\ \hline
$\Phi\Pi$ & $0$ & $1$ & $-1$ & $0$ & $-1$ & $0$ & $-1$ & $-2$
\\ \hline
\end{tabular}

\vspace{.2cm}{\footnotesize {\bf Table }
Ultraviolet, infrared dimensions and Faddeev--Popov
charges.}
\end{center}
The stability of the theory can be checked by perturbing the classical
action~$\Sm$ with an integrated functional~$\S_c$, which, according to
the quantum action principle~\cite{qap}, is the most general one having
$d_{uv}\leq 4$ and $d_{ir}\geq 4$. After imposing the identities defining the
theory on the perturbed action~$\Sm+\e\S_c$, we will show
that the perturbation~$\S_c$ can be reabsorbed in~$\Sm$ by a
number of redefinitions of the fields and parameters, which counts the
renormalization constants.

As in the massless case~\cite{psbk}, because of the linearity of the
gauge condition~\equ{gaugecond}, the functional $\S_c$ does not
depend on the multiplier $b(x)$. Moreover, it contains the
antighost~$\cb(x)$
and the source~$\O(x)$ only through the combination
\eq
\Oh^{a\m}=\O^{a\m}+\pa^\m\cb^a\ .
\eqn{comb}
In the Landau gauge, the two additional symmetries of the classical
action~\equ{bps} and~\equ{w} imply that the ghost~$c(x)$ and the
source~$\s(x)$
appear only differentiated. In addition, since the mass to the YM fields is
provided by the spontaneous symmetry breaking in the direction of the
external field~$\t(x)$, the set of constraints on~$\S_c$ is completed by
\eq
\left.\fud{\S_c}{\t(x)}\right|_{\psi=0}=0\ \ ,\ \ \psi(x)=\mbox{all fields}\ ,
\eqn{ssb}
and by the shift equation controlling the dependence of the classical
action~$\Sm$ upon the massive parameter~$m^2$
\eq
\Lp \pad{}{m^2} - \intx \fud{}{\t(x)} \Rp \Sm =0\ .
\eqn{shiftcond}
Finally, the Slavnov identity~\equ{slavnov} imposed on
the perturbed action, at first order in the infinitesimal parameter~$\e$
translates into the following Slavnov condition
\eq
\BS\S_c =0\ ,
\eqn{bs}
where $\BS$ is the linearized Slavnov operator
\eq
\BS = \intx \Lp
\fud{\Sm}{\Oh^{a\m}}\fud{}{A^a_\m}
+\fud{\Sm}{A^a_\m}\fud{}{\Oh^{a\m}}
+\fud{\Sm}{L^a}\fud{}{c^a}
+\fud{\Sm}{c^a}\fud{}{L^a}
+\th\fud{}{\s} \Rp \ ,
\eqn{linslavnov}
which, by effect of~\equ{slavnov}, is nilpotent
\eq
(\BS)^2=0\ .
\eqn{nil}
The Slavnov condition~\equ{bs} is easily solved once we remark
that the only difference with respect to the massless case is the
presence of the external sources~$\s$ and~$\th$, which appear in the
Slavnov operator~\equ{linslavnov} as a BRS doublet, and consequently non
altering the cohomological structure of the theory~\cite{psbk}.
Therefore, the solution of the condition~\equ{bs} can depend on the new
fields~$\s(x)$ and $\th(x)$ only through a trivial cocycle.

The most general solution, satisfying the whole set of constraints, is
\eq
\S_c = Z_g\intx F^2 + \BS \intx \Lp
Z_A \Oh^{a\m}A^a_\m + Z_c L^ac^a + Z_m\s A^2 \Rp\ ,
\eqn{counter}
where the four constants~$(Z_g,Z_A,Z_c,Z_m)$ correspond respectively to
renormalizations of the gauge coupling constant, the YM field, the ghost
field and the parameter~$m^2$. We remark that in the Landau gauge, as a
consequence of the symmetries~\equ{bps} and~\equ{w}, the ghost field and
the mass parameter do not renormalize
\eq
Z_c = Z_m = 0\ \ \ \mbox{(Landau gauge)}.
\eqn{nonren}

For what concerns the presence of anomalies, again the fact that the
external fields~$\s(x)$ and~$\th(x)$ are BRS doublets, insures that,
algebraically, the only anomaly is the ABJ one, which has
a vanishing coefficient, since all the fields are in the adjoint
representation of the gauge group. One may easily verify also the absence
of infrared anomalies, \ie of counterterms having infrared
dimension~$d_{ir}<4$.

In the CF model, the correlation functions of physical observables depend
upon the gauge parameter~$\a$, and it has been argued that
the $\a-$independence could be recovered
when~$m^2=0$~\cite{cf1,peri}. In our case, the
introduction of the mass parameter does not change the dependence of the
classical action on the gauge parameter~$\a$, nor the fact that it doesn't
renormalize. Hence, the usual proof~\cite{ps} of gauge
independence of the theory based on the ``extended'' BRS symmetry, holds
true. The argument of~\cite{ps} consists in extending the classical
identity
\eq
\pad{\S_m}{\a}=\BS \intx \half \cb^ab^a
\eqn{ext}
to all orders of perturbation theory
\eq
\pad{\G}{\a} = {\cal B}_\G (\D\cdot\G)\ ,
\eqn{qext}
where $\G$ is the quantum vertex functional and~$\D\cdot\G$ is a quantum
insertion, by exploiting the non-renormalization properties of the
parameter~$\a$, or, in other words, the linearity of the gauge
condition~\equ{gaugecond}. The quantum relation~\equ{qext} states the
non-physical character of the gauge parameter, from which follows that
the correlators of the physical observables are independent from~$\a$.
The nature of the parameter~$m^2$ is different, since even classically
it does not satisfy an identity analogous to~\equ{ext}. 
We can only say that the physical observables do not depend from the
shifted external field $\hat\tau (x)$, since the cohomology of the linearized
Slavnov operator is independent from it. Nevertheless a $m^2$ dependence of the
Green functions of the physical operators cannot be excluded; indeed 
Eq.~\equ{shiftcond}, 
which is to become the Callan-Symanzik equation in the quantized theory,
controls the mass dependence of amplitudes as due to radiative
corrections. 

\section{Conclusions}

We provided the YM fields of a mass~$m^2$, through  the spontaneous
symmetry breaking in the direction of an external source, in the
framework of a local and renormalizable theory. In absence of unitarity,
the mass we introduced must be considered an infrared regulator, and
not a physical mass. Unlike what happens in the CF model, we are able to
write a true Slavnov identity, with a linear gauge choice. The physical
observables do not depend on the gauge parameter~$\a$. 
Moreover, in the Landau gauge the
mass~$m^2$ does not renormalize. 
A precise analysis needs a 
choice of a renormalization procedure; we can only 
comment that in the dimensional scheme the counterterms are polynomials in any 
dimensional parameter and therefore the effective Lagrangian possesses a smooth 
zero mass limit. For this reason, and since the external field whose shift 
gives mass to the gauge fields does not appear in the cohomology space, we 
expect that the zero mass limit of the massive model (modulo gauge fixing 
delicacies) should go smoothly to the usual massless Yang-Mills 
theory~\cite{bm}.

\vspace{2cm}

{\bf Acknowledgments} We would like to thank O.Piguet, M.Porrati and S.Wolf
for useful discussions and V.Periwal, for some clarifications concerning
Ref.~\cite{peri}.


\begin{thebibliography}{999}
\bibitem{cf}      G.Curci and R.Ferrari, \nc{32}{76}{151}
\bibitem{pvn}     J.de Boer, K.Skenderis, P.van Nieuwenhuizen and
                  A.Waldron, \pl{B367}{96}{175}
\bibitem{peri}    V.Periwal, ``Infrared regularization of non-abelian gauge
                  theories'', Princeton preprint, hep-th/9509084; and
                  ``Unitary theory of massive non-abelian vector bosons''
                  Princeton preprint, hep-th/9509085.
\bibitem{delsor}  F.Delduc and S.P.Sorella, \pl{B231}{89}{408}
\bibitem{cf1}     G.Curci and R.Ferrari, \nc{35}{76}{1}
\bibitem{ojima}    I.Ojima, \zp{C13}{82}{173}
\bibitem{bau}     L.Baulieu, \prep{129}{85}{1}
\bibitem{del}     R.Delbourgo, S.Twisk and G.Thompson, \ijmp{A3}{88}{435}
\bibitem{bps}    A.Blasi, O.Piguet and S.P.Sorella, \np {B356}{91}{154}
\bibitem{psbk}   O.Piguet and S.P.Sorella, {\it Algebraic
                 Renormalization}, Lectures notes in Physics, vol. m28,
                 Springer Verlag 1995.
\bibitem{qap}    J.H.Lowenstein, \pr{D4}{71}{2281} \cmp{24}{71}{1}
                 Y.M.P.Lam, \pr{D6}{72}{2145} \pr{D7}{73}{2943}
                 T.E.Clark and J.H.Lowenstein, \np{B113}{76}{109}
\bibitem{ps}     O.Piguet and K.Sibold, \np{B253}{84}{517}
\bibitem{bm}     A.Blasi and N.Maggiore, in preparation.
\end{thebibliography}
\end{document}